\documentclass[pra,twocolumn,english,showpacs,nofootinbib]{revtex4-1}
\usepackage{amssymb}
\usepackage{color}
\usepackage{graphicx}
\usepackage{bbold}
\usepackage{bm}
\usepackage{amsmath}
\usepackage{amsfonts}
\usepackage{amssymb}
\usepackage{braket}
\usepackage{units}
\usepackage{csquotes}
\usepackage{upgreek}
\usepackage{xcolor}
\usepackage{fancyhdr}
\begin{document}

\title{Correlated photon-pair emission from a cw-pumped Fabry-Perot microcavity}

\author{Thorsten F. Langerfeld, Hendrik M. Meyer, Michael K\"ohl}

\affiliation{Physikalisches Institut, Universit\"at Bonn, Wegelerstrasse 8, 53115 Bonn, Germany}

\begin{abstract}
We study a dispersion-compensated high-finesse optical Fabry-Perot microcavity under high-intensity cw pumping. The Kerr non-linearity in the optical coatings causes a spontaneous four-wave mixing process, which leads to the emission of time-correlated photon pairs. The photon frequencies are shifted  by $\pm 1$ free spectral range relative to the pump frequency. This setup allows for constructing a photon-pair source with precisely adjustable frequency difference between  the emitted photons, which may have applications in quantum communication.
\end{abstract}

\maketitle

\section{Introduction}

The generation of correlated photons is an important milestone in fundamental test of quantum mechanics \cite{Hong1987,Bouwmeester1997} and in the quest to interconnect remote quantum systems with the goal of creating quantum networks \cite{Kimble2008}. For the latter, both bandwidth and frequencies of the photons have to be tailored to the physical properties of the network nodes, for which photon pair sources with a broad wavelength tuning and narrow-band linewidth are required. Moreover, for interfacing  emitters with different resonance wavelengths, a tuneable photon pair source, which can bridge a large and tuneable wavelength gap, is desirable. This is of particular importance for solid state quantum emitters, networks of which \cite{Delteil2017} are often challenged by inhomogeneities in sample fabrication, and for hybrid quantum networks between distinct emitters \cite{Meyer2015}.

The most common method of photon pair production is spontaneous parametric downconversion (SPDC). However, also other techniques such as spontaneous four-wave mixing (SFWM) have been demonstrated. The different approaches have their own merits and challenges. SPDC sources with short crystals and SFWM sources, for example in optical fibers \cite{Fiorentino2002,Fan2005}, employ high-intensity, short-pulse pump lasers propagating through non-linear optical media. The simplicity of these schemes results from the weak requirements regarding phase matching and leads to correlated photon pair emission in a broad bandwidth of several THz.  For applications in which the correlated photons are to be interfaced with solid state or atomic samples. However, the broad bandwidth is a significant hurdle since the bandwidth of absorption is in the range of 10--1000 MHz. Tight constraints regarding phase matching often go hand-in-hand with restrictions regarding the emission wavelengths and/or efficiencies. To this end, narrow-band photon-pair sources have been developed. However, this comes at the experimental expense of controlling the phase matching condition for the desired wavelength(s) and combining this with spectral filtering techniques. Additionally, the non-linear medium often is placed in an optical cavity to to enhance the nonlinearity. For an SPDC source this has been demonstrated to achieve bandwidth/frequency matching with a trapped single ion \cite{Haase2009}. Moreover, work on microcavities in silicon have demonstrated photon pair generation using SFWM with typical cavity radii of a few ten micrometers and Q-Factors up to $5\times 10^5$. For a recent compilation of data see \cite{Savanier2016}.

In this manuscript, we demonstrate a particularly simple setup for the creation of correlated photon pairs:  We pump a Fabry-Perot cavity consisting of two dielectric mirrors with light from a single-frequency  diode laser and utilize the optical Kerr effect  in predominantly one layer of the dielectric mirror coating for spontaneous four-wave mixing. Previously, nonlinear effects in optical coatings have been discussed in the context of short-pulse lasers \cite{Fedulova2016} but here we demonstrate the first results with a continuous-wave laser. The four-wave mixing process absorbs two photons from the pump light field at frequency $\omega_0$ and produces photon pairs at frequencies $\omega_{n,\pm}=\omega_0\pm n\cdot \omega_{FSR}$, where $\omega_{FSR}=\frac{\pi c}{L}$ denotes the free spectral range of the Fabry-Perot cavity of length $L$, $c$ is the speed of light, and $n$ is an integer. The spectral bandwidth of the emitted photon pairs is given by the cavity bandwidth (the decay rate of the intracavity electric field) $\kappa = \omega_{FSR}/(2F)$ where $F$ denotes the finesse of the resonator. The generation of single photons through spontaneous four-wave mixing in nonlinear media is triggered by vacuum fluctuations of the unoccupied cavity modes $\omega_0\pm \omega_{FSR}$. We can consider the thermal occupation of these modes as negligible even for our room-temperature cavity since $(\omega_0\pm \omega_{FSR})/(k_BT) \sim 60$.

\section{Experimental setup}

\begin{figure}[htbp]
\centering
\includegraphics[width=0.95\columnwidth]{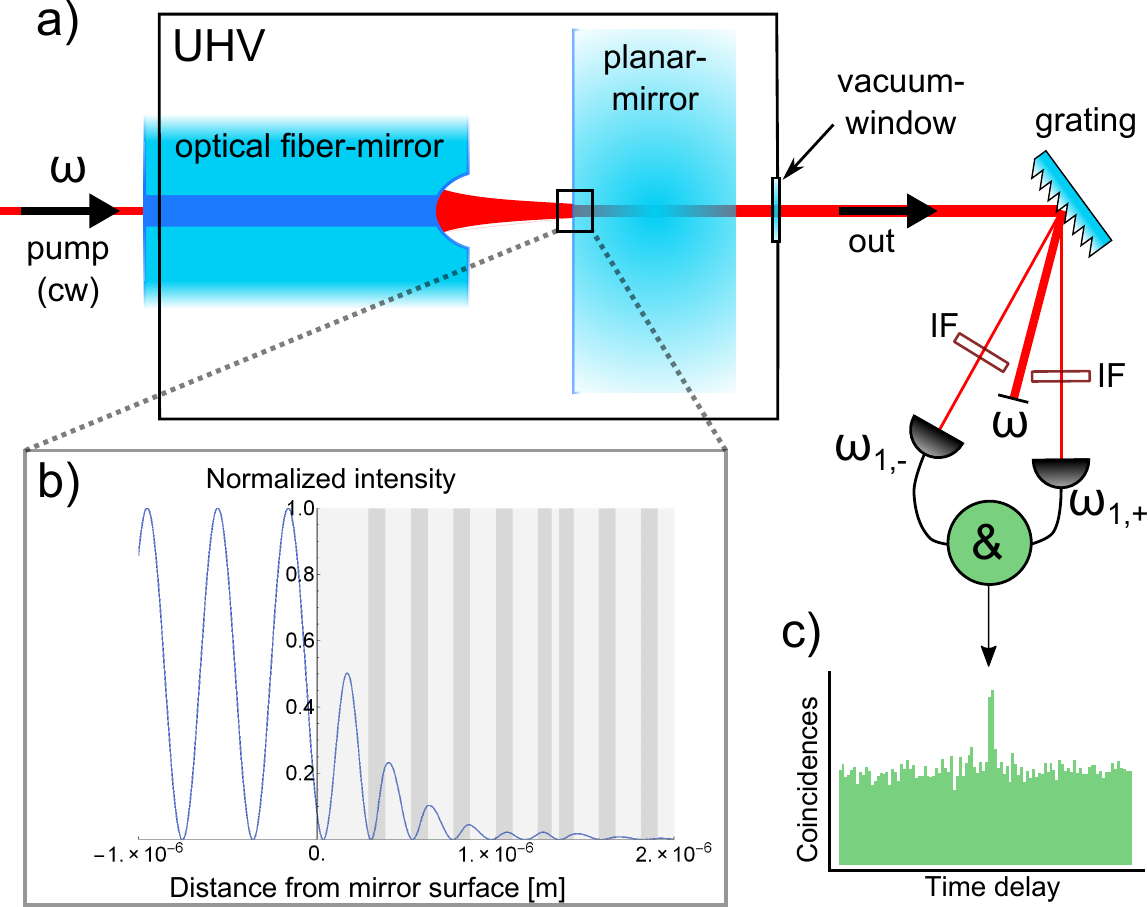}
\caption{(a) Setup of the Fabry-Perot cavity consisting of a machined and coated end facet of an optical fiber and a conventional mirror in an ultrahigh vacuum chamber. The outcoupled light is dispersed by a grating spectrometer and photons of the two longitudinal modes $\omega_{1,\pm}$ adjacent to the pump mode $\omega$ are guided to single photon counters. Further filtering is performed by narrow-linewidth interference filters (IF). (b) Simulation of the laser field in the optical coating. Light gray: SiO$_2$, dark gray Nb$_2$O$_5$. Since the Kerr coefficient for Nb$_2$O$_5$ is more than one order of magnitude larger than that of SiO$_2$, the main effect results from the second layer of the coating. (c) Sample of pair correlations. }
\label{fig1}
\end{figure}

Our Fabry-Perot cavity is composed of a micromachined, and coated endfacet of an optical fiber as one mirror\cite{Hunger2010,Muller2010,Steiner2013,Brandstaetter2013,Takahashi2014,Uphoff2015,Gallego2016,Janitz2017} and a conventional planar mirror with identical coating as the second mirror (see Figure \ref{fig1}|a). The length of the cavity is $L=30\,\mu$m, corresponding to a free-spectral range of $\omega_{FSR}=2\pi\times 4.5\,$THz, and has a finesse of $F=\pi/(T+L)=4500\pm 60$ with a nominal mirror transmission of $T=100$\,ppm and intracavity losses of $L=600$\,ppm per mirror.  The radius of curvature of the fiber mirror is $R=170\,\mu$m, giving rise to a $1/e^2$-beam radius on the planar mirror of $w_0=4\,\mu$m. This small mode waist enhances the desired non-linear effects. Both mirrors have the same dielectric coating featuring $\sim 40$ layers, alternatingly of SiO$_2$ and Nb$_2$O$_5$ as the low- and high-index material, respectively. In Figure 1b, we show a numerical simulation of the optical fields inside the coating of the mirrors of the Fabry-Perot resonator. From the simulation, we confirm that the main nonlinear effect of the optical coating results from the second layer, since the non-linear refractive indices of the materials of the coatings are $n_2$(SiO$_2)=3\times 10^{-20}$\,m$^2$/W and $n_2$(Nb$_5$O$_2)=2\times 10^{-19}$\,m$^2$/W \cite{Dimitrov1996}.

We have precisely determined the zero-dispersion frequency of the cavity by measuring the frequency difference between the cavity mode and the higher (+1) and lower (--1) longitudinal modes for several pump frequencies. In Figure 2a we show the frequency difference between these modes as a function of the center wavelength. For the optimized pump wavelength of 379.3\,THz  we have experimentally verified that the free spectral ranges with $n=\pm 1$ agree to within 170\,MHz, which is a small fraction of the cavity linewidth of $\kappa=2\pi\times 1$\,GHz. The strong pumping of the cavity with intracavity intensities of  $\lesssim 10^{11}$\,W/m$^2$  leads to significant thermal effects in the coatings. We observe the characteristic \cite{Dube1996} bistable cavity line shape, see Figure 2b. The different line shapes result from the thermal expansion of the coating shifting the resonance frequency either in the direction of the cavity scan or opposite. In order to eliminate the thermal effects, we lock the cavity in our experiment to the pump laser using a Pound-Drever-Hall locking scheme and we observe a  power variation of 10\%. During the course of the experiment, we have observed a gradual degradation of the cavity finesse from its initial value of $10^4$ to the value of 4500 quoted above and at which the measurements were performed. Because surface contamination of the cavity mirrors within a UHV environment is unlikely, we attribute the increased losses to weak optical damage from the high intensity laser irradiation.

We monitor the output of the Fabry-Perot cavity using a home-built grating spectrometer and a pair of single photon counting modules (SPCM) protected from stray pump light by additional dielectric line filters, see Figure 1a. Both line filters have a linewidth of 3nm and transmit light at the respective detection wavelength (781nm, 799nm) while providing a suppression of the pump stray field on the SPCM of at least five orders of magnitude. The grating spectrometer has a resolution of $\lambda/\delta\lambda=6000$ and the measured quantum efficiencies of the two beam paths are 36\% and 40\%. The timing jitter of the SPCM is 350\,ps. We record the photon counter signals using a time-to-digital converter with timing resolution of 40\,ps for several hours and subsequently perform a correlation analysis with adjustable bandwidth.

\begin{figure}[!htbp]
\centering
\includegraphics[width=\linewidth]{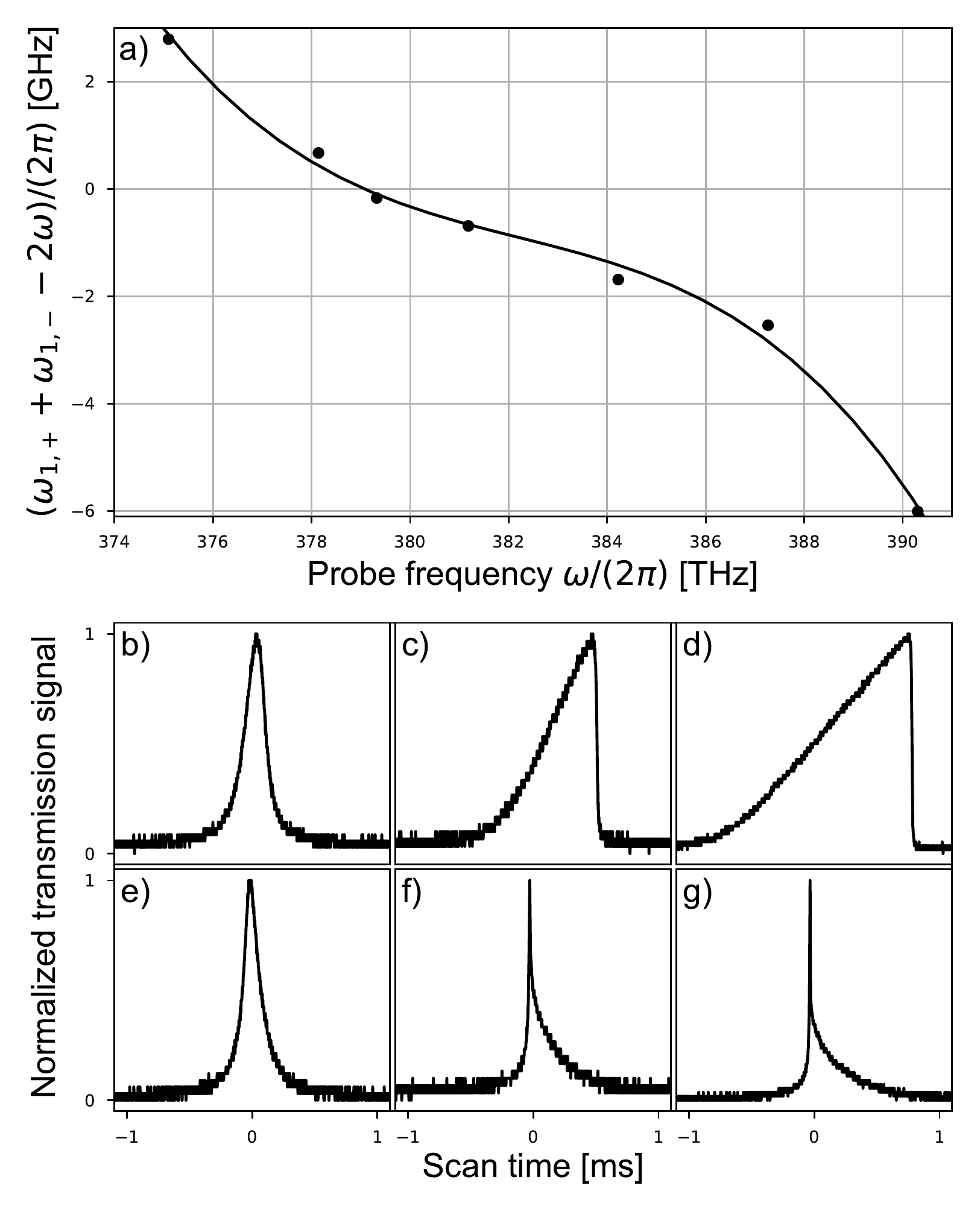}
\caption{ (a) Determination of the zero-dispersion of the cavity. Shown is the frequency difference between the higher and lower free-spectral range of the cavity. The zero-crossing denotes the pump frequency for which the cavity dispersion disappears. (b-g) Optical bistability by absorption in the mirrors. The top row shows the scan from blue- to red-detuning and the bottom row shows the scan from red- to blue-detuning for the following intracavity powers: b/c: 700\,mW, d/e: 7\,W, and f/g: 14\,W.}
\label{fig1a}
\end{figure}

\section{Results}
In Figure 3 we show a sample of the two-photon coincidence signal for an intracavity power of 19\,W.  The signal comprises of a uniform background of spurious coincidences and an excess peak near zero time delay between the counters. The small delay is caused by the slightly different optical beam paths and lengths of the electric cabling between detector and counter. We fit the correlation signal with a Lorentzian function on top of a constant background in order to determine both the rate and bandwidth of the correlated photon pairs.

It has been pointed out that both spontaneous four-wave mixing and Raman scattering can generate single photons. In spontaneous Raman scattering  the photons are created by scattering from phonons in the optical material. Hence the photon emission rate in Raman scattering depends linearly on the pump field intensity and, moreover, Raman scattering creates an excess of (lower energy) idler photons over (higher energy) signal photons. Additionally, spontaneous Raman scattering does not contribute to real coincidences at a fixed time difference but produces accidental coincidences at arbitrary time difference. Since the cavity is pumped with a continuous-wave field this only leads to a constant offset of the measured coincidences, while real coincidences created by photon pairs always arrive within the cavity ring-down time-constant near zero time delay.

In Figure 4a we show the measured coincidence rate $R_{12}$ versus optical pump power, which is extracted from the sum of the measured coincidences within the Lorentzian peak after subtraction of the offset. The data show a quadratic dependence on the pump-power as expected for spontaneous four-wave mixing. The solid line is a quadratic fit $R_{12}=\Gamma\cdot P^2$ to the data. Here $P$ refers to the intracavity power, which is connected to time-averaged intensity of the cavity standing wave field by $I=\frac{2  F P}{\pi^2 w_0^2}$.

We compute the expected flux of photon pairs from SFWM from the optical cavity as $\Gamma=\frac{\omega_{FSR}F}{\pi} \left[k\int n_2(x) I(x) dx\right]^2=0.6$\,W$^{-2}$s$^{-1}$ \cite{Lamprecht1996}. The quantities $n_2(x)$ and $I(x)$ are the nonlinear refractive index and the light intensity, respectively, as a function of the penetration depth into the coating, and $k=2\pi/\lambda$ is the wave vector of the light. The integral of the  Kerr coupling constant runs over both mirror coatings.  In order to compare the numerical analysis to the experimental data, we take into account that the outcoupling efficiency of a photon from the cavity into the spectrometer direction is given by $T/(2T+2L)=0.07$, the transmission of the optical setup is 90\% and the detector efficiencies are quoted above. Resultingly, we obtain an expected flux of $\Gamma^\prime=3\times 10^{-4}$ \,W$^{-2}$s$^{-1}$. Experimentally, we find a $\Gamma_\text{exp}=1.2(1)\times 10^{-4}$ \,W$^{-2}$s$^{-1}$. Since our calculation does not include the dielectric relaxation time which would reduce the dielectric response at optical frequencies \cite{Boettcher1996}, we consider this a fair agreement between experiment and theory. Additionally, in Figure 4b we show the measured bandwidth $\tau_c$ of the correlation signal. This bandwidth agrees excellently with the inverse cavity linewidth. On both single channels we find a linear dependence of the count rate on the incoupled power, which we fit by $R_i = \gamma_i\cdot P$ giving $\gamma_1 = 275(5)~\text{W}^{-1}\text{s}^{-1}$ and $\gamma_2 = 257(4) ~\text{W}^{-1}\text{s}^{-1}$ respectively. The background rate are 1-2 orders of magnitude above the detector dark count rates and we suspect that the background results from Raman scattering. From these single channel count rates we expect a maximal ratio of real to accidental coincidences of $\frac{2}{\pi \tau_c}\left(\frac{\Gamma_{\text{exp}}}{\gamma_1 \gamma_2}\right) = 0.99 \pm 0.24$, which is in agreement with the ratio of 0.82 extracted from Figure 3. We have extracted the formular by assuming a Cauchy distribution of the correlated photon pairs of width $\tau_c$ and a binning time much shorter than the correlation time, which is well fulfilled in our experiment. 

\begin{figure}[!htbp]
\centering
\includegraphics[width=\linewidth]{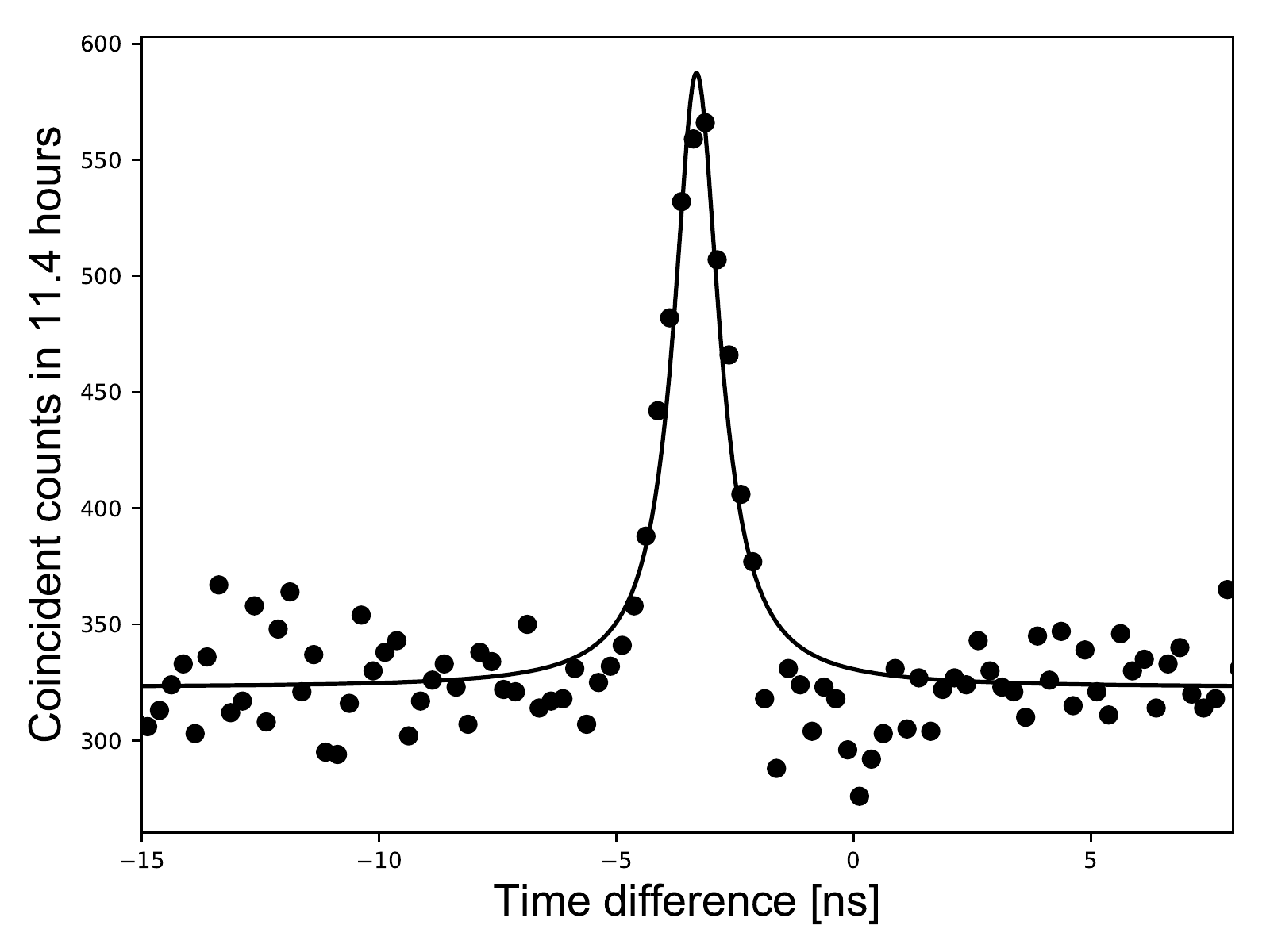}
\caption{Sample trace of detected photon pair correlations (binning $\Delta \tau = 250~$ps). Data are taken for an intracavity power of 19\,W.}
\label{fig3}
\end{figure}

\begin{figure}[!htbp]
\centering
\includegraphics[width=\linewidth]{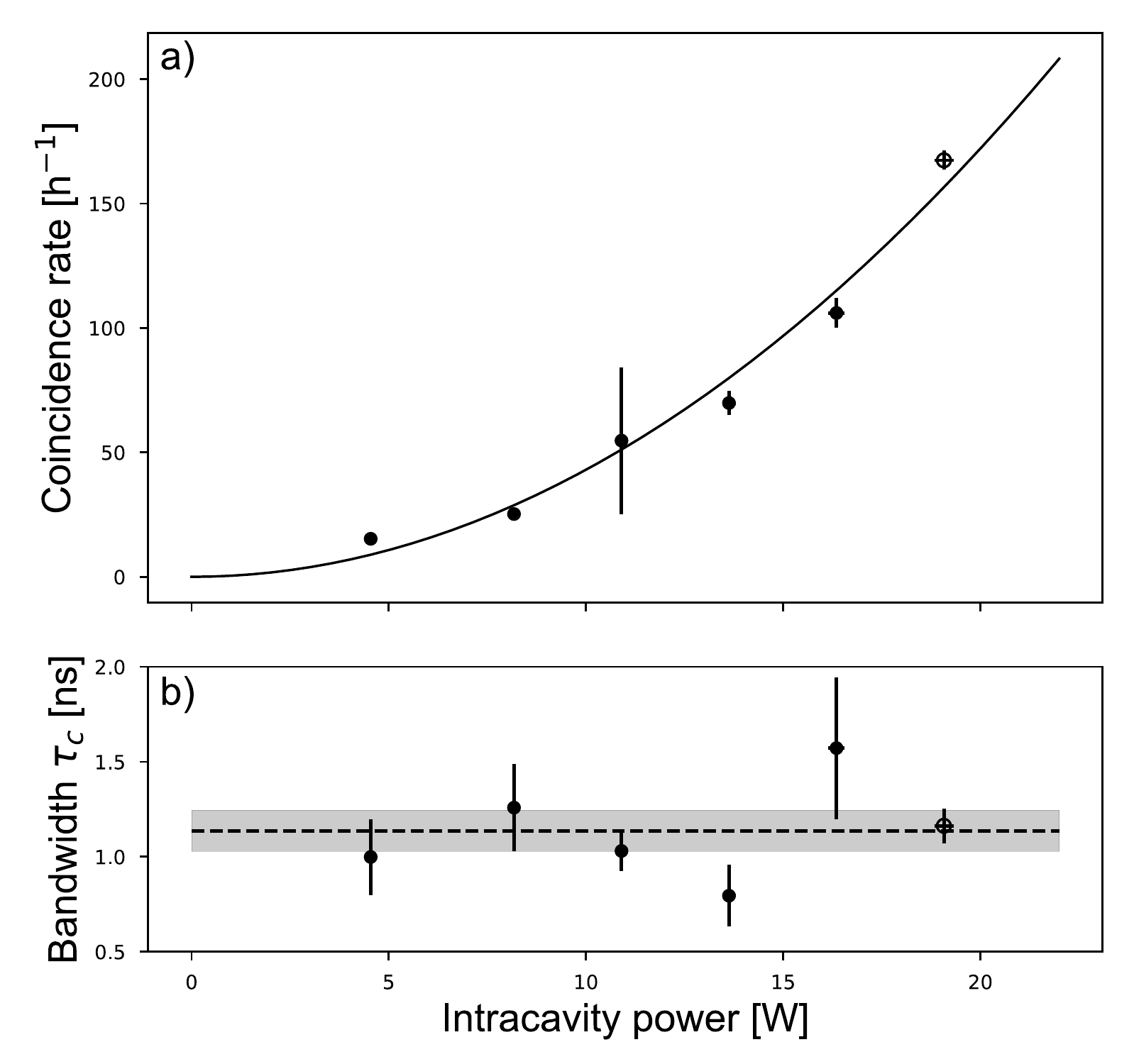}
\caption{(a) Pair correlations vs. pump intensity. The solid line is a quadratic fit to the data to extract the pair generation efficiency. (b) Width of the pair correlation peak vs. pump intensity. The dashed line indicates the lifetime of the photons in the cavity. The hollow points in both plots signal the data set shown in Figure 3.}
\label{fig4}
\end{figure}

In conclusion, we have observed the generation of correlated photon pairs by spontaneous four-wave mixing from a pumped high-finesse Fabry-Perot cavity utilizing the intrinsic non-linearity of the dielectric mirrors. The ease of the experimental setup, e.g. avoiding a phase matching condition by employing a sub-wavelength thick nonlinear medium, and the principal tunability of the wavelengths and bandwidths of the created photon pair make the scheme an attractive candidate for a photon-pair source with application in hybrid quantum systems \cite{Meyer2015} in which wavelength has to be bridged between dissimilar systems. There are two ways in which a wavelength adjustment can be considered: (1) using higher-order free spectral ranges, i.e. $n\geq 2$. This would require to shift the center frequency of the cavity in order to ensure dispersion compensation. (2) change the cavity length. This would facilitate to work at the same center frequency but with adjustable free spectral range. Both ways are experimentally feasible. The pair production rate can be enhanced in the future by optimizing mirror coatings.  

We thank I. Carusotto, M. Fleischhauer, T. Gross, and M. Schultze for discussions and the Alexander-von-Humboldt Stiftung, Deutsche Forschungsgemeinschaft (SFB/TR 185) and the Bonn-Cologne Graduate School of Physics and Astronomy (BCGS) for support.

\end{document}